\documentclass[aps,prd,twocolumn,preprintnumbers,superscriptaddress,nofootinbib]{revtex4-1}
\usepackage{graphicx}
\usepackage{epstopdf}
\usepackage{amsmath}
\usepackage{amsfonts}
\usepackage{amssymb}
\usepackage{appendix}
\usepackage{enumerate}
\usepackage{natbib}
\usepackage{comment}
\usepackage{bbold}
\usepackage[shortlabels]{enumitem}
\usepackage{color}
\usepackage{slashed}
\usepackage{subfigure}
\usepackage{setspace}
\usepackage{footnote}
\usepackage{lipsum}
\usepackage{multirow}
\usepackage{float}
\usepackage[colorlinks = true,
            linkcolor = blue,
            urlcolor  = blue,
            citecolor = blue,
            anchorcolor = blue]{hyperref}
\usepackage[capitalize]{cleveref}
\usepackage{braket}
\usepackage[compat=1.1.0]{tikz-feynman}
\usepackage{multirow}
\usepackage{physics}
\usepackage{feynmp-auto}
\usepackage[normalem]{ulem}
\usepackage{url}
\usepackage{units}
\usepackage[normalem]{ulem}

\newcommand{\be}{\begin{eqnarray}}
\newcommand{\ee}{\end{eqnarray}}
\newcommand{\ba} {\begin{equation}\begin{aligned}}
\newcommand{\ea} {\end{aligned}\end{equation}}
\newcommand{\bg} {\begin{equation}\begin{gathered}}
\newcommand{\eg} {\end{gathered}\end{equation}}

\newcommand{\beq}{\begin{equation}}
\newcommand{\eeq}{\end{equation}}

\usepackage{tikz,xcolor,hyperref}

\definecolor{lime}{HTML}{A6CE39}
\DeclareRobustCommand{\orcidicon}{\hspace{-1mm}
	\begin{tikzpicture}
		\draw[lime, fill=lime] (0,0) 
		circle [radius=0.12] 
		node[white] {{\fontfamily{qag}\selectfont \tiny \,ID}};
		\draw[white, fill=white] (-0.0525,0.095) 
		circle [radius=0.007];
	\end{tikzpicture}
	\hspace{-3mm}
}

\foreach \x in {A, ..., Z}{\expandafter\xdef\csname orcid\x\endcsname{\noexpand\href{https://orcid.org/\csname orcidauthor\x\endcsname}
		{\noexpand\orcidicon}}
}

\begin{document}

\title{Searching for Axions from Atmospheric Kaon Decays}

\author{Tousif Raza\orcidA}
\email{tousif.m.raza@gmail.com}

\affiliation{Department of Physics, Oklahoma State University, Stillwater, OK 74078, USA}

\date{\today}

\begin{abstract}
Cosmic-ray interactions in the Earth's atmosphere produce copious secondary mesons, providing a natural laboratory for searches for new physics. We investigate the atmospheric production of axions through the rare kaon decay $K^{+}\rightarrow\pi^{+}a$. If sufficiently long-lived, these axions can reach large underground
detectors such as Super-Kamiokande and IceCube and decay through the diphoton channel, $a\rightarrow\gamma\gamma$. Using
data from Super-Kamiokande and IceCube, we derive constraints on the axion decay constant $f_a$ for masses up to $m_a\sim 350~\mathrm{MeV}$. We demonstrate that atmospheric kaon decays provide a novel probe of axion parameter space, yielding constraints complementary to existing bounds.

\end{abstract}

\maketitle

\section{Introduction} {\label{sec:intro}}
Experimental searches constrain the neutron electric dipole moment to
$d_n\lesssim 2\times10^{-26}\,e\,\mathrm{cm}$
~\cite{2016PhRvL.116p1601G,Abel:2020pzs}. Within QCD, this stringent
bound implies that the effective vacuum angle governing strong CP
violation must satisfy $|\bar{\theta}|\lesssim10^{-10}$. The absence of
a natural explanation for such a small value constitutes the strong CP
problem. The Peccei--Quinn mechanism provides an elegant solution by
introducing a global $U(1)_{\mathrm{PQ}}$ symmetry that is spontaneously
broken, thereby promoting the effective QCD vacuum angle to a dynamical
degree of freedom. The resulting field relaxes to the CP-conserving
minimum of its QCD-induced potential
~\cite{PhysRevLett.38.1440,PhysRevD.16.1791}. The
pseudo-Nambu--Goldstone boson associated with this symmetry breaking is
the axion~\cite{PhysRevLett.40.223,PhysRevLett.40.279}.

More generally, axion-like particles (ALPs) constitute a broad class of pseudoscalar bosons that appear in numerous extensions of the Standard Model. They may play important roles in cosmology, including driving
inflation~\cite{Freese:1990rb,Freese:2004un,Silverstein:2008sg,Kaloper:2008fb,Kim:2004rp}, constituting dark
matter~\cite{Preskill:1982cy,Abbott:1982af,Dine:1982ah}, or contributing to dark energy~\cite{Frieman:1995pm,Kaloper:2005aj,Kaloper:2008fb,
Nomura:2000yk,Ibe:2018ffn,Choi:2019jck,Choi:2021aze}. String compactifications provide a well-motivated theoretical setting for axions and ALPs, which can arise from higher-dimensional antisymmetric
tensor fields after compactification~\cite{Witten:1984dg,Svrcek:2006yi,
Arvanitaki:2009fg}.

These theoretical motivations have stimulated extensive efforts to investigate the phenomenology of axions, ALPs, and other weakly coupled light particles~\cite{Ringwald:2012hr,Essig:2013lka,Graham:2015ouw, Marsh:2017hbv,Hook:2019qoh,Choi:2020rgn}. A variety of accelerator experiments and proposed facilities probe such particles through complementary production and detection
channels~\cite{BDX:2016akw,LDMX:2018cma,PIENU:2021clt,
PIONEER:2022yag,Goudzovski:2022vbt,NA64:2020qwq}.
Beam-dump experiments provide a well-established setting for searches for axions and ALPs. In this work, we instead consider their production in the Earth's atmosphere. The nearly isotropic cosmic-ray flux, composed predominantly of protons, continuously interacts with atmospheric nuclei, producing an abundant population of secondary particles, including mesons and leptons. The broad energy spectrum and
continuous flux of cosmic rays therefore make the atmosphere a natural
fixed-target environment for producing weakly coupled particles beyond
the Standard Model.

We investigate the production of heavy axions through the
flavor-changing neutral-current decay $K^{+}\to\pi^{+}a$ of charged
kaons produced in cosmic-ray-induced atmospheric showers. If
sufficiently long-lived, these axions can propagate to large
underground detectors and decay within their instrumented volumes.
Using Super-Kamiokande and IceCube data, we search for their decays
through the diphoton channel, $a\to\gamma\gamma$, and derive
constraints on the axion parameter space for
$m_a\lesssim350~\mathrm{MeV}$, covering nearly the entire mass range
kinematically accessible in $K^{+}\to\pi^{+}a$.

The remainder of this article is organized as follows.
\Cref{sec:axion} reviews the effective-field-theory framework employed
in our analysis, while \cref{sec:ALP-production} describes atmospheric
axion production through charged-kaon decays.
\Cref{sec:ALP-detection} presents our analysis of Super-Kamiokande and
IceCube data and the resulting constraints.
Finally, \cref{sec:conclusion} concludes the paper.
\section{Axion EFT}
\label{sec:axion}

We briefly review the effective field theory (EFT) framework relevant
to our axion analysis~\cite{Dolan:2017osp,Ertas:2020xcc,Bauer:2021wjo}.
The relevant interactions with the SM gauge fields are
\begin{equation}
\begin{split}
\mathcal{L} ={}&
\frac{1}{2}\partial^\mu a\,\partial_\mu a
-\frac{1}{2}m_a^2a^2
+c_{GG}\frac{\alpha_s}{4\pi}\frac{a}{f}
G_{\mu\nu}\widetilde{G}^{\mu\nu}
\\
&+
c_{WW}\frac{\alpha_w}{4\pi}\frac{a}{f}
W_{\mu\nu}\widetilde{W}^{\mu\nu}
+c_{BB}\frac{\alpha_Y}{4\pi}\frac{a}{f}
B_{\mu\nu}\widetilde{B}^{\mu\nu}\,.
\end{split}
\label{eq:alp_eft}
\end{equation}
where $a$ and $m_a$ denote the axion field and mass, respectively, and
$c_{GG}$, $c_{WW}$, and $c_{BB}$ are dimensionless Wilson coefficients. Here, $\alpha_s=g_s^2/(4\pi)$, $\alpha_w=g^2/(4\pi)$, and
$\alpha_Y=g'^2/(4\pi)$ denote the couplings associated with the
$SU(3)_c$, $SU(2)_W$, and $U(1)_Y$ gauge sectors, respectively. We adopt the convention

\begin{equation}
f_a=\frac{f}{2c_{GG}}\,.
\label{eq:fa_definition}
\end{equation}

We consider two representative benchmark scenarios. In the
gluon-dominance scenario,
\begin{equation}
c_{GG}\neq 0,\qquad c_{WW}=c_{BB}=0\,,
\end{equation}
whereas in the co-dominance scenario,
\begin{equation}
c_{GG}=c_{WW}=c_{BB}\neq 0\,.
\end{equation}
These benchmarks illustrate the effects of gluonic and electroweak
interactions on axion production and decay, and hence on the sensitivity
of atmospheric searches.

\section{Axion Production in the Atmosphere}

\label{sec:ALP-production}

\begin{figure*}[htp]

\setkeys{Gin}{width=0.45\linewidth}
    {\includegraphics{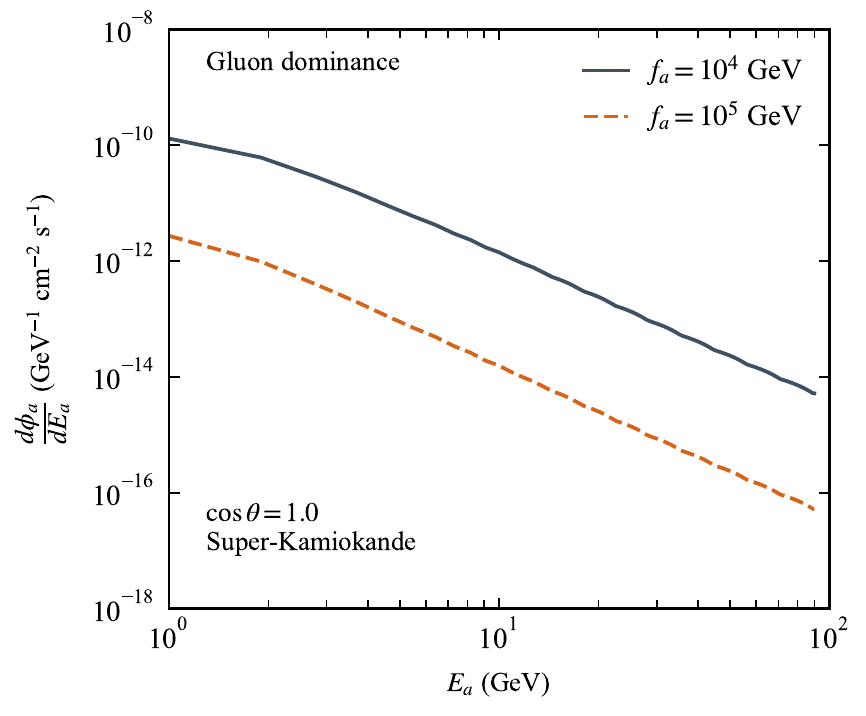}}
    {\includegraphics{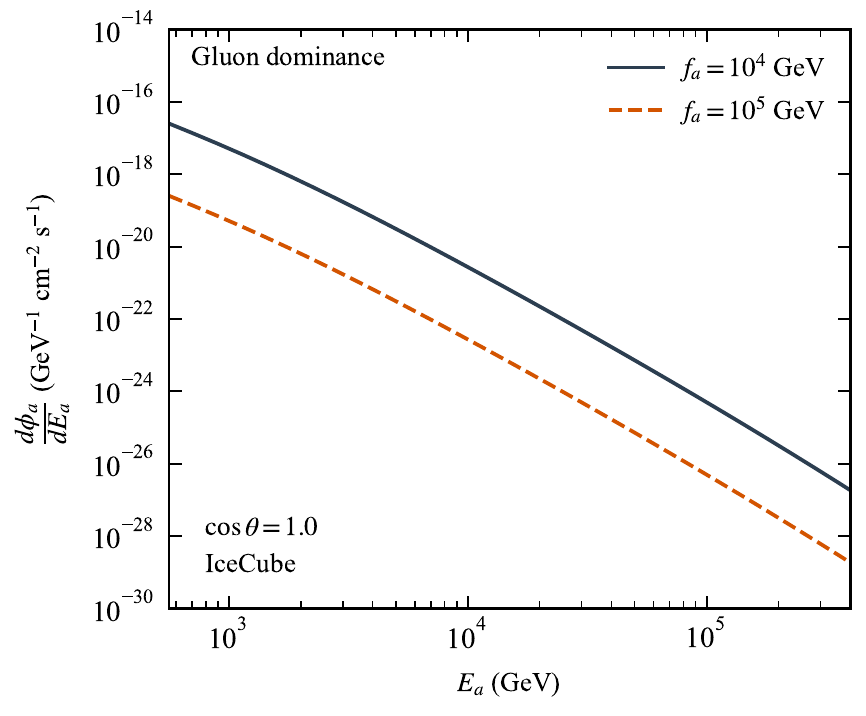}}

\caption{
Atmospheric axion flux from the decay $K^{+}\to\pi^{+}a$, evaluated
at the Earth's surface for $\cos\theta=1$ after integrating over the
atmospheric production height. Results are shown for
$m_a=160~\mathrm{MeV}$ in the axion-energy ranges relevant to
Super-Kamiokande (\textbf{left}) and IceCube (\textbf{right}) in the
gluon-dominance scenario.
}
\label{fig:flux_plots}
\end{figure*}

Flavor-changing neutral-current (FCNC) transitions provide a natural
mechanism for producing axions through the rare kaon decay
\begin{equation}
K^{+}\rightarrow\pi^{+}a\,.
\label{eq:kaon_decay}
\end{equation}
Although this production channel has been extensively studied in laboratory and beam-dump settings~\cite{NA62:2023olg,NA62:2021zjw,NA62:2025upx}, here we consider a naturally occurring realization in the Earth's
atmosphere. High-energy cosmic rays continuously interact with
atmospheric nuclei, producing secondary hadrons, including a substantial
flux of charged kaons, which can subsequently decay into axions through
\cref{eq:kaon_decay}.

To model the atmospheric kaon flux, we use the Matrix Cascade
Equations (MCEq) numerical code~\cite{Fedynitch:2015zma}.
In our simulations, we adopt the Hillas--Gaisser model for the primary
cosmic-ray spectrum~\cite{Gaisser:2011klf}, the NRLMSISE-00
atmospheric model~\cite{https://doi.org/10.1029/2002JA009430}, and SIBYLL-2.3c for hadronic
interactions~\cite{Fedynitch:2018cbl}. The MCEq framework also allows
for alternative choices of the primary cosmic-ray
spectrum~\cite{2012PhRvD..86k4024F} and hadronic interaction models,
including QGSJET-III~\cite{Ostapchenko:2024myl},
DPMJET-III~\cite{Roesler:2000he,Fedynitch:2015kcn}, and
EPOS-LHC~\cite{Pierog:2013ria}. For the purposes of the present
analysis, we do not explicitly include the associated model
dependence.

Given the atmospheric kaon flux, the differential contribution to the axion flux from a production point located a distance $l$ from the detector is obtained as~\cite{Gondolo:1995fq,Arguelles:2019ziu}
\begin{equation}
\frac{d\phi_a}
{dE_a\,d\cos\theta\,dl}
=
\int dE_K\,
\frac{1}{l_K}
\frac{d\phi_K}{dE_K\,d\cos\theta}
\frac{dn}{dE_a}(K\rightarrow a)
e^{-l/l_a}\,.
\label{eq:cascade}
\end{equation}
Here, $d\phi_K/(dE_K\,d\cos\theta)$ denotes the parent charged-kaon
flux, while $l_K=\gamma_K\beta_K c\tau_K$ is the kaon decay length in
the laboratory frame. The exponential factor $e^{-l/l_a}$ gives the
probability that an axion produced in the atmosphere survives
propagation over a distance $l$ before reaching the detector, where
$l_a=\gamma_a\beta_a c\tau_a$ is the axion decay length in the
laboratory frame. Here, $\gamma_i=E_i/m_i$ and $\beta_i=p_i/E_i$, with
$i\in\{K,a\}$. The total axion
flux at the detector is obtained by integrating
\cref{eq:cascade} over the atmospheric production region.

For a given zenith angle $\theta$, the propagation distance from an
atmospheric production point at altitude $h$ to the detector is fixed
by the geometry of the Earth and satisfies
\begin{equation}
l^2+2Rl\cos\theta+
\left[R^2-(R+h)^2\right]=0\,,
\label{eq:atmospheric_distance}
\end{equation}
where $R$ denotes the Earth's radius and $h$ the altitude of the production point in the adopted atmospheric model.
This geometric relation allows the atmospheric production height to be
mapped onto the propagation distance entering the survival factor in
\cref{eq:cascade}.

The energy spectrum of axions produced in the two-body decay
$K^{+}\rightarrow\pi^{+}a$ is given by
\begin{equation}
\frac{dn}{dE_a}
=
\frac{\mathrm{BR}(K^{+}\rightarrow\pi^{+}a)}
     {\Gamma(K^{+}\rightarrow\pi^{+}a)}
\frac{d\Gamma(K^{+}\rightarrow\pi^{+}a)}{dE_a}\,.
\label{eq:axion_energy_spectrum}
\end{equation}
Within the effective-field-theory framework, the branching ratio for
this rare kaon decay can be normalized to the experimentally well-measured
decay $K_S\rightarrow\pi^{+}\pi^{-}$ as~\cite{Bauer:2021wjo}
\begin{equation}
\frac{\mathrm{BR}(K^{+}\rightarrow\pi^{+}a)}
     {\mathrm{BR}(K_S\rightarrow\pi^{+}\pi^{-})}
=
\frac{\tau_{K^{+}}}{\tau_{K_S}}
\frac{f_\pi^2}{8f_a^2}
\left|
1+\frac{c_{uu}+c_{dd}}{c_{GG}}
\right|^2\,.
\label{eq:kaon_decay_br}
\end{equation}
Here, $\tau_{K^{+}}$ and $\tau_{K_S}$ denote the proper lifetimes of the
kaon and the short-lived neutral kaon, respectively, while
$f_\pi$ is the pion decay constant. We adopt
$\mathrm{BR}(K_S\rightarrow\pi^{+}\pi^{-})\simeq 69.2\%$ as the
experimental normalization~\cite{ParticleDataGroup:2024cfk}. Throughout our numerical analysis, we assume
$(c_{uu}+c_{dd})/c_{GG}=1$.
Alternatively, neglecting corrections of
$\mathcal{O}(m_\pi^2/m_K^2)$ and $\mathcal{O}(m_a^2/m_K^2)$, one may adopt
$(c_{uu}+c_{dd})/c_{GG}=0$ in the branching ratio, as considered in
Ref.~\cite{Ema:2023tjg}. Relative to our benchmark choice, this
alternative reduces $\mathrm{BR}(K^{+}\rightarrow\pi^{+}a)$, and hence
the expected signal yield, by a factor of four. Since the branching
ratio scales as $f_a^{-2}$, the corresponding sensitivity to $1/f_a$
is reduced by a factor of two.

For an unpolarized two-body decay, the normalized axion-energy
distribution in the laboratory frame is
\begin{equation}
\frac{1}{\Gamma}
\frac{d\Gamma}{dE_a}
=
\frac{1}{p_K}
\frac{1}{
\sqrt{
\lambda\left(
1,\frac{m_\pi^2}{m_K^2},\frac{m_a^2}{m_K^2}
\right)
}
}\,,
\label{eq:normalized_spectrum}
\end{equation}
where $p_K$ denotes the kaon momentum and
$\lambda$ is the K\"all\'en function,
\begin{equation}
\lambda(x,y,z)
=
x^2+y^2+z^2-2xy-2xz-2yz\,.
\label{eq:kallen}
\end{equation}

The two-body decay kinematics also determine the range of parent-kaon
energies that contributes to a given axion energy. In the kaon rest
frame, the axion is monoenergetic, with energy
\begin{equation}
E_{a}^\ast
=
\frac{m_K^2-m_\pi^2+m_a^2}{2m_K}\,.
\label{eq:Eamax}
\end{equation}
After boosting to the laboratory frame, the allowed axion energy for a
kaon with energy $E_K$ and momentum $p_K$ is bounded by
\begin{equation}
\begin{split}
\gamma_K E_{a}^\ast
-\frac{p_K}{2}
\sqrt{\lambda\left(
1,\frac{m_\pi^2}{m_K^2},\frac{m_a^2}{m_K^2}
\right)}
&\leq E_a \leq \\
\gamma_K E_{a}^\ast
+\frac{p_K}{2}
\sqrt{\lambda\left(
1,\frac{m_\pi^2}{m_K^2},\frac{m_a^2}{m_K^2}
\right)}\,
\end{split}
\label{eq:Ea_bounds}
\end{equation}
where $\gamma_K=E_K/m_K$ is the Lorentz factor of the parent kaon.
For the flux calculation it is useful to invert these relations and determine the range of parent-kaon energies that can produce an axion with a specified laboratory-frame energy $E_a$. In the highly relativistic limit, $E_K\gg m_K$, the corresponding bounds are
\begin{equation} \begin{split} \frac{m_K E_a} {E_{a}^\ast +\dfrac{m_K}{2} \sqrt{\lambda\left( 1,\frac{m_\pi^2}{m_K^2},\frac{m_a^2}{m_K^2} \right)}} &\leq E_K \leq \\[2pt] \frac{m_K E_a} {E_{a}^\ast -\dfrac{m_K}{2} \sqrt{\lambda\left( 1,\frac{m_\pi^2}{m_K^2},\frac{m_a^2}{m_K^2} \right)}} \end{split} \label{eq:EK_bounds} \end{equation}
These kinematic bounds determine the range of the atmospheric
parent kaon spectrum that contributes to the axion flux at a given
axion energy and mass.
\section{ Analysis and Results}

\label{sec:ALP-detection}

\begin{figure*}[htp]
\setkeys{Gin}{width=0.45\linewidth}
{\includegraphics{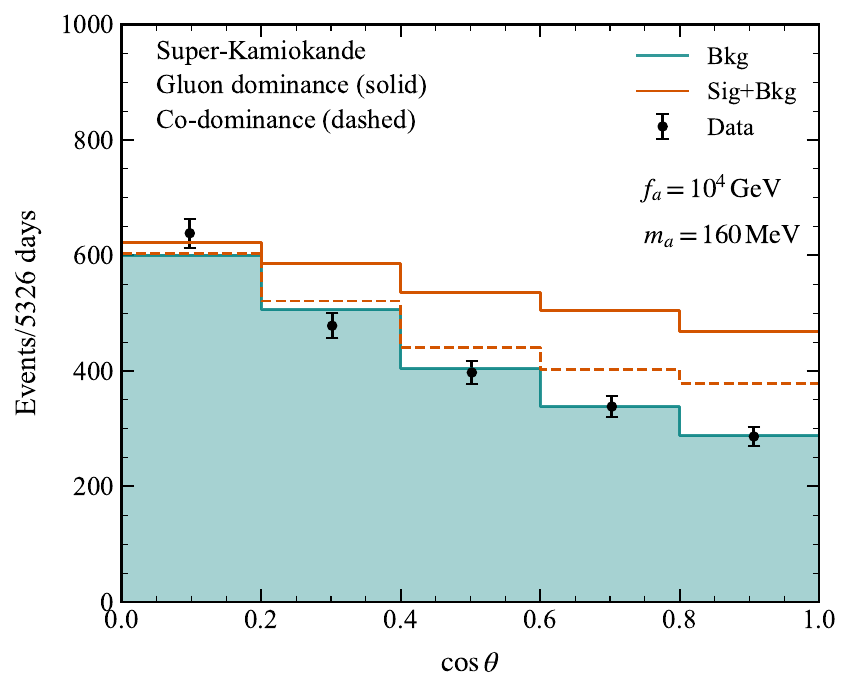}}
{\includegraphics{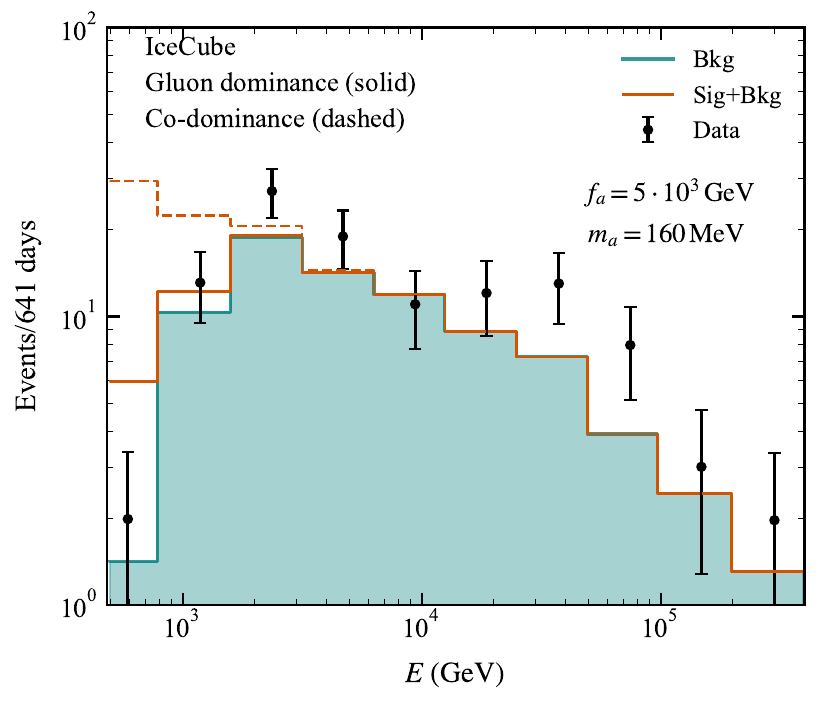}}
\caption{Number of expected axion events from parent kaon decays in
the gluon-dominance and co-dominance scenarios.
\textbf{Left panel:} Expected axion events at Super-Kamiokande as a
function of the zenith angle.
\textbf{Right panel:} Expected axion events at IceCube as a function
of energy. In both panels, the backgrounds are shown as teal histograms~\cite{Super-Kamiokande:2017yvm,IceCube:2014rwe}.}
\label{fig:evts_benchmark_Super-K}
\end{figure*}

Axions produced through atmospheric kaon decays can propagate to underground detectors and decay within their fiducial volumes, provided that they are sufficiently long-lived. The probability for an axion to reach a detector is governed by its laboratory-frame decay length, which
depends on both its proper lifetime and Lorentz boost. In the mass range $m_a<3m_\pi$, and assuming that the diphoton channel dominates, the axion lifetime can be represented as
\begin{equation}
\tau_a=\Gamma_a^{-1}
=
\frac{256\pi^3 f_a^2}
{\alpha^2m_a^3|c_{\mathrm{eff}}|^2}\,,
\label{eq:alp_lifetime}
\end{equation}
where $\alpha$ is the electromagnetic fine-structure constant and
$c_{\mathrm{eff}}$ denotes the effective axion-photon coupling
coefficient,
\begin{equation}
c_{\mathrm{eff}}
=
c_{\gamma\gamma}
-
\left(
\frac{5}{3}
+
\frac{m_\pi^2}{m_\pi^2-m_a^2}
\frac{m_d-m_u}{m_d+m_u}
\right)\,.
\label{eq:ceff}
\end{equation}
%
The scaling $\tau_a\propto f_a^2/m_a^3$ implies that, at fixed $m_a$, increasing $f_a$ generally increases the lifetime and allows a larger fraction of the atmospheric axion flux to reach the detector. At fixed $f_a$, heavier axions generally have shorter lifetimes and are therefore more likely to decay before reaching the detector.

In our analysis, we set
$c_{\gamma\gamma}=0$ in the gluon-dominance scenario and
$c_{\gamma\gamma}=2$ in the co-dominance scenario following the treatment adopted in
Ref.~\cite{Ema:2023tjg}. 
For fixed $m_a$ and $f_a$,
the different effective couplings in the two benchmark scenarios lead
to different axion lifetimes and consequently different decay lengths. The laboratory-frame decay length is further enhanced by the Lorentz boost, an effect that is particularly important for energetic axions produced in cosmic-ray-induced kaon decays.

To determine the expected signal at Super-Kamiokande and IceCube, we
compute the number of axions that reach the detector and subsequently
decay within its volume. The expected number of signal events during
the data-taking interval $\Delta T$ is given by

\begin{equation}
\begin{split}
N &= \mathrm{BR}(a\to\gamma\gamma) \\
&\times \int d\cos\theta \int dE_a\,
\epsilon\,\Delta T\,A_{\mathrm{decay}}^{\mathrm{eff}}\frac{d\phi_a}
{dE_a\,d\cos\theta}
\end{split}
\label{eq:event_rate}
\end{equation}

Here, $\mathrm{BR}(a\to\gamma\gamma)$ denotes the axion branching ratio
into two photons, which are reconstructed as $e$-like events.
For the numerical results presented below, we adopt a reference efficiency of $\epsilon=0.75$ for Super-Kamiokande, following
Ref.~\cite{Arguelles:2019ziu}, and assume unit efficiency for IceCube.
The resulting IceCube sensitivity should therefore be interpreted as
an idealized estimate.  
The effective decay area $A_{\mathrm{decay}}^{\mathrm{eff}}$ accounts
for the probability that an axion entering the detector decays along
its trajectory within the detector volume. It is obtained by
integrating over the detector surface perpendicular to the incident
axion direction, weighted by the corresponding decay probability,
\begin{equation}
A_{\mathrm{decay}}^{\mathrm{eff}}
=
\int dS_{\perp}
\left(1-e^{-\Delta l/l_a}\right)\,,
\label{eq:effective_decay_area}
\end{equation}
where $\Delta l$ denotes the length of the axion trajectory contained
within the detector. For Super-Kamiokande, we consider the detector as a cylinder with height
$H=0.04~\mathrm{km}$ and radius $R=0.02~\mathrm{km}$ and use
reconstructed zenith angles in the range $0\leq\cos\theta\leq1$. For
IceCube, we adopt $H=1~\mathrm{km}$ and
$R=1/\sqrt{\pi}~\mathrm{km}$, with reconstructed zenith angles in the
range $0.2\leq\cos\theta\leq1$, following
Ref.~\cite{Arguelles:2019ziu}.

Using the atmospheric axion flux derived in
\cref{sec:ALP-production}, together with the detector response described
above, we calculate the expected signal distributions at
Super-Kamiokande and IceCube. In
\cref{fig:evts_benchmark_Super-K}, we present the expected event
distributions for a representative axion mass of
$m_a=160~\mathrm{MeV}$ in the gluon-dominance and co-dominance
benchmark scenarios. The left panel shows the expected number of events
at Super-Kamiokande as a function of the zenith angle, while the right
panel shows the expected number of events at IceCube as a function of
energy.

The expected event rate exhibits a characteristic dependence on the
axion decay constant. At small $f_a$, axion production through kaon
decays is enhanced; however, the corresponding axion lifetime is
shorter, and the flux reaching the detector is exponentially suppressed
by the survival probability $e^{-l/l_a}$. At large $f_a$, the longer
lifetime allows a larger fraction of the produced axions to reach the
detector, while the production rate is suppressed. The detectable event
rate is therefore governed by the interplay among axion production,
propagation through the atmosphere, and the probability of decay within
the detector volume. For a given energy, heavier axions also have
smaller Lorentz boosts and generally shorter decay lengths, further
affecting the expected event rate.
\begin{figure*}[t]
\setkeys{Gin}{width=0.45\linewidth}
    {\includegraphics{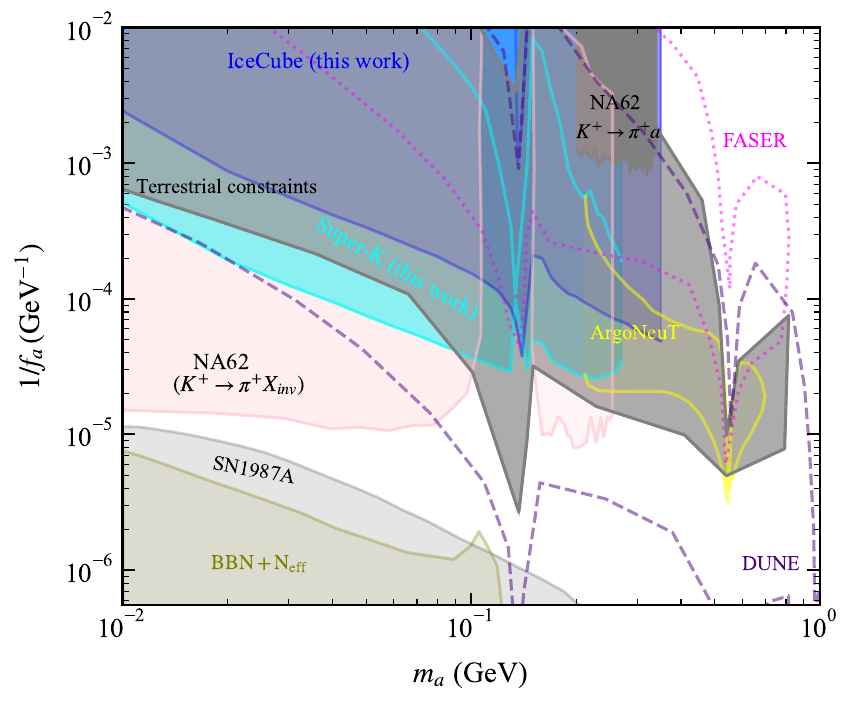}}
    {\includegraphics{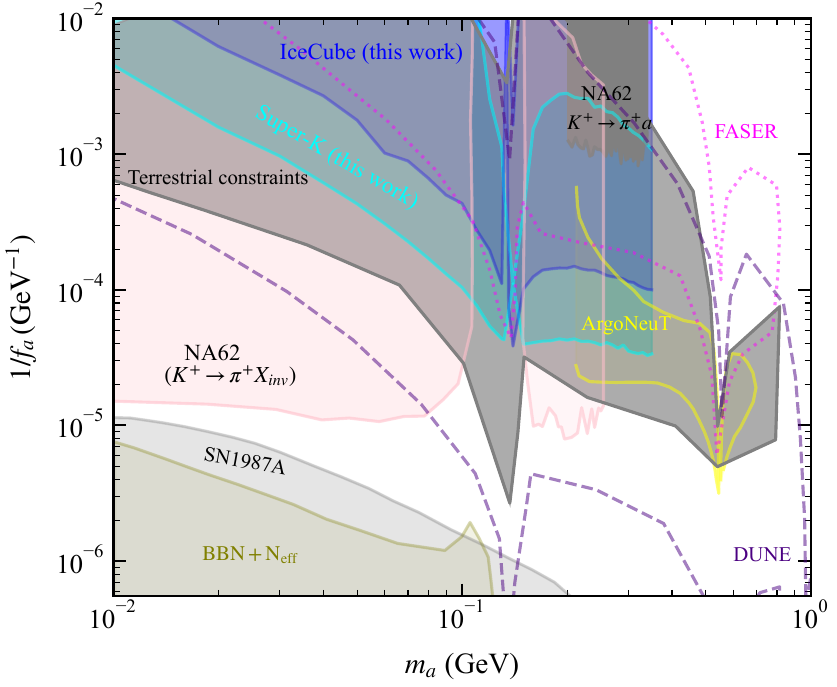}}

\caption{The $90\%$ C.L. limits on the inverse axion decay constant,
$1/f_a$, derived from Super-Kamiokande and IceCube data.
\textbf{Left panel:} Gluon-dominance scenario.
\textbf{Right panel:} Co-dominance scenario.
In both panels, the Super-Kamiokande limits derived from atmospheric
kaon decays are shown in cyan, while the IceCube limits are shown in
blue. We compare our results with existing terrestrial constraints
from E949~\cite{BNL-E949:2009dza}, electron beam-dump
experiments~\cite{Dobrich:2015jyk,Dolan:2017osp},
CHARM~\cite{CHARM:1985anb}, NuCal~\cite{Blumlein:1990ay}, LHC dijet
searches~\cite{Gavela:2019cmq}, ArgoNeuT~\cite{ArgoNeuT:2022mrm}, and
NA62~\cite{NA62:2021zjw,NA62:2023olg}. We also show astrophysical
constraints from SN1987A~\cite{Chang:2018rso} and
cosmological constraints from Ref.~\cite{Depta:2020wmr}. The projected
sensitivities of DUNE~\cite{Kelly:2020dda} and
FASER~\cite{FASER:2018eoc} are indicated by dashed and dotted lines,
respectively. For additional projected sensitivities, see
Refs.~\cite{Blinov:2021say,SHiP:2018xqw,
Coloma:2023oxx,Ema:2025bww}.}
\label{fig:exclusion_limits}
\end{figure*}

To derive exclusion limits from the predicted event distributions, we
use data from Super-Kamiokande~\cite{Super-Kamiokande:2017yvm} and
IceCube~\cite{IceCube:2014rwe} and construct the Poisson
likelihood-ratio test statistic
\begin{equation}
\chi^2
=
2\sum_k
\left[
N_{o,k}
\ln\left(
\frac{N_{o,k}}{N_{s,k}+N_{b,k}}
\right)
-N_{o,k}
+N_{s,k}
+N_{b,k}
\right]\,,
\label{eq:chi2_stat}
\end{equation}

where $N_{s,k}$, $N_{b,k}$, and $N_{o,k}$ denote the expected signal,
background, and observed numbers of events in the $k$th bin,
respectively. For Super-Kamiokande, the sum runs over the angular bins,
whereas for IceCube it runs over the energy bins. 
For each fixed axion mass $m_a$, we treat $1/f_a$ as the single parameter of interest. The corresponding 90\% C.L. exclusion
boundaries are determined using the criterion
$
\Delta\chi^2(1/f_a;m_a)
=
\chi^2(1/f_a,m_a)-\chi^2_{\min}(m_a)
=
2.71
$.

In \cref{fig:exclusion_limits}, we show the 90\% C.L. constraints on the inverse axion decay constant, $1/f_a$, derived from Super-Kamiokande and IceCube data. The left and right panels correspond to the gluon-dominance and co-dominance benchmark scenarios, respectively. At small $1/f_a$, production through $K^+\to\pi^+a$ is
suppressed, while the long axion lifetime reduces the probability of decay within the detector. Conversely, at large $1/f_a$, the production rate is enhanced, but the shorter lifetime causes a larger fraction of
the axions to decay before reaching the detector. The strongest sensitivity therefore occurs at intermediate values of $1/f_a$, for which an appreciable atmospheric axion flux reaches the detector and subsequently decays within its fiducial volume. The cyan and blue shaded
regions are excluded by the Super-Kamiokande and IceCube data, respectively.

We find that, over most of the kinematically accessible mass range,
Super-Kamiokande provides stronger constraints than IceCube. This
difference is primarily attributable to the lower axion-energy range
covered by the published Super-Kamiokande data used in our analysis,
$E_a\simeq 1$--$90~\mathrm{GeV}$, where the atmospheric parent-kaon
flux and consequently the axion flux are comparatively larger. The
lower boundaries of the Super-Kamiokande exclusion regions reach
values of $1/f_a$ of order
$10^{-5}~\mathrm{GeV}^{-1}$\footnote{Atmospheric constraints on
muonphilic axion-like particles produced through three-body
charged-pion decays were previously derived in
Ref.~\cite{Cheung:2022umw}. Following Ref.~\cite{Eberhart:2025lyu},
these limits can be mapped onto the $(m_a,1/f_a)$ plane using
$C_\ell/\Lambda=g_{a\ell}/m_a$ and $\Lambda\simeq4\pi f_a$. Under this
coupling convention, our results substantially improve on the previous
atmospheric constraints in the kinematically accessible mass range,
$m_a<m_{\pi^\pm}-m_{\mu^\pm}$.}. The published IceCube data used in
our analysis cover substantially higher energies, approximately
$562~\mathrm{GeV}$--$446~\mathrm{TeV}$, where the atmospheric
parent-kaon flux is smaller. The resulting IceCube sensitivity is
therefore generally weaker, reaching values of $1/f_a$ of order
$10^{-4}~\mathrm{GeV}^{-1}$ over part of the mass range
$m_a\simeq 100$--$350~\mathrm{MeV}$. Nevertheless, the larger Lorentz
boosts of axions in the IceCube energy range increase their decay
lengths in the laboratory frame. IceCube consequently retains
sensitivity at comparatively larger values of $1/f_a$, for which
lower-energy axions would predominantly decay before reaching
Super-Kamiokande. The two experiments therefore probe complementary
regions of the axion parameter space, with sensitivities comparable to
existing terrestrial bounds~\cite{BNL-E949:2009dza,Dobrich:2015jyk,
CHARM:1985anb,Blumlein:1990ay,Gavela:2019cmq,ArgoNeuT:2022mrm,
NA62:2021zjw,NA62:2023olg}.

\begin{figure*}[htp]

\setkeys{Gin}{width=0.45\linewidth}
    {\includegraphics{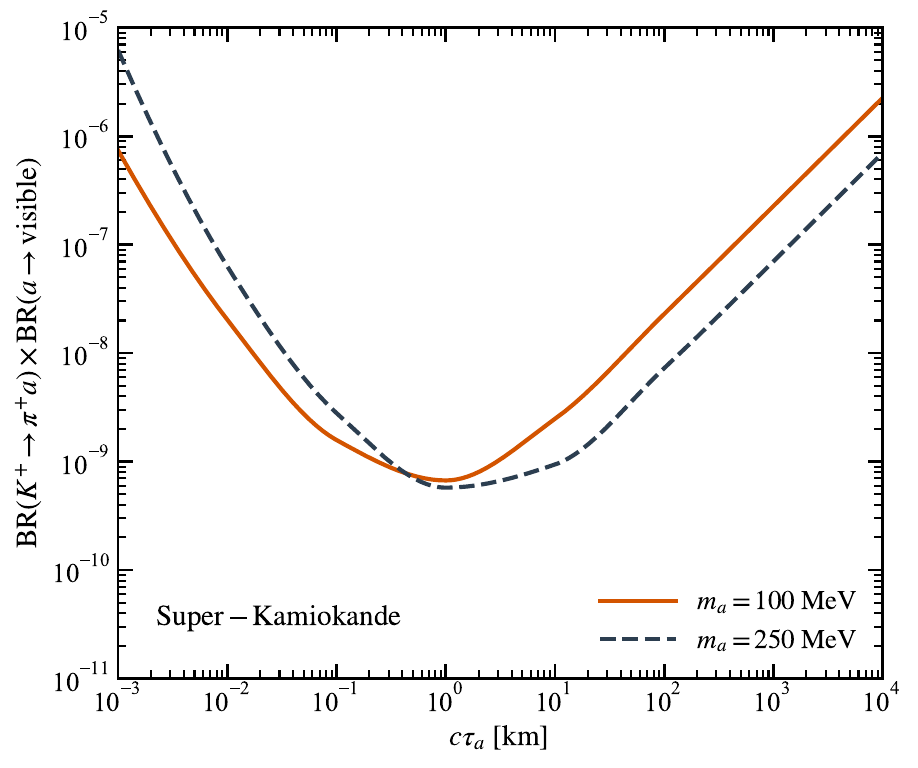}}
    {\includegraphics{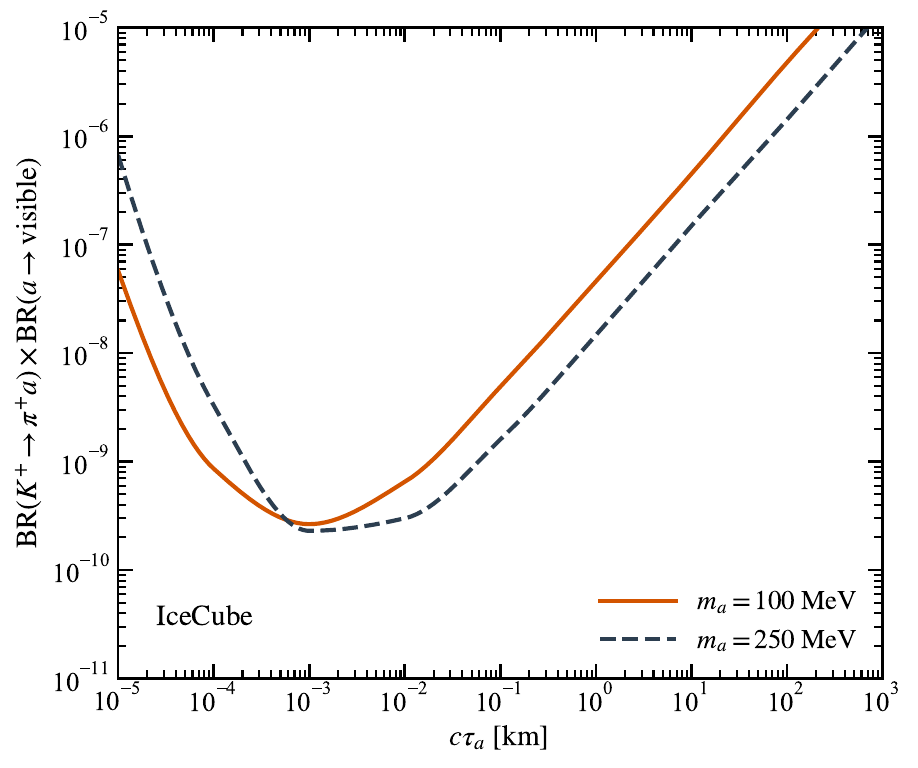}}

\caption{
The $90\%$ C.L. upper limits on the branching-ratio product
$\mathrm{BR}(K^{+}\to\pi^{+}a)\times
\mathrm{BR}(a\to\mathrm{visible})$ as a function of the axion proper
decay length for $m_a=100$ and $250~\mathrm{MeV}$. The Super-Kamiokande and IceCube results are shown in the \textbf{left} and \textbf{right} panels, respectively.}
\label{fig:BR_plots}
\end{figure*}

Alongside the exclusion limits on the axion decay constant, we performed
a separate model-independent analysis in which we constrain the
branching-ratio product
$\mathrm{BR}(K^{+}\to\pi^{+}a)
\times\mathrm{BR}(a\to\mathrm{visible})$
as a function of the axion proper decay length $c\tau_a$.

In \cref{fig:BR_plots}, we present the 90\% C.L. upper limits obtained
from Super-Kamiokande and IceCube on
$\mathrm{BR}(K^{+}\to\pi^{+}a)
\times\mathrm{BR}(a\to\mathrm{visible})$
as a function of $c\tau_a$ for two representative masses,
$m_a=100~\mathrm{MeV}$ and $250~\mathrm{MeV}$. For each fixed choice of $(m_a,c\tau_a)$, we derive the 90\% C.L. upper limit on this branching-ratio product, treating it as the single parameter of interest.
The strongest Super-Kamiokande and IceCube limits on the
branching-ratio product reach values of order $10^{-10}$. Their optimal sensitivities occur at proper decay lengths of approximately
$c\tau_a\sim1$--$3~\mathrm{km}$ and
$c\tau_a\sim10^{-3}$--$10^{-2}~\mathrm{km}$, respectively. This model-independent presentation enables a direct comparison with
searches for the two-body decay $K^{+}\to\pi^{+}X$, followed by the visible decay of a long-lived particle $X$. In the relevant mass and lifetime regions, the sensitivity obtained
from atmospheric kaon decays does not exceed the existing limits
reported by NA62~\cite{NA62:2023olg,NA62:2025upx} and
MicroBooNE~\cite{MicroBooNE:2021sov}.
\section{Conclusion}
\label{sec:conclusion}

We have investigated the atmospheric production of axions through the
rare decay $K^{+}\rightarrow\pi^{+}a$ of charged kaons generated in
cosmic-ray-induced air showers. Using the MCEq framework, we modeled the
atmospheric kaon flux and derived the corresponding axion flux,
accounting for the two-body decay kinematics, production-altitude
distribution, and survival probability. For masses below the kinematic
threshold, $m_a<m_K-m_\pi$, sufficiently long-lived axions can reach
underground neutrino detectors and decay within their fiducial volumes.
We focused on the diphoton channel, $a\rightarrow\gamma\gamma$, as the
observable signature.

Using published Super-Kamiokande and IceCube data, we performed binned
likelihood analyses of the zenith-angle distribution at Super-Kamiokande
and the reconstructed-energy distribution at IceCube. For the
gluon-dominance and co-dominance benchmark scenarios, we derived
constraints on the axion decay constant $f_a$ across the kinematically
accessible mass range. Super-Kamiokande benefits from the larger
atmospheric kaon flux at lower energies, whereas the greater Lorentz
boosts at IceCube increase the laboratory-frame axion decay length,
allowing sensitivity to shorter proper lifetimes.

We also derived limits on
$\mathrm{BR}(K^{+}\rightarrow\pi^{+}a)
\times\mathrm{BR}(a\rightarrow\mathrm{visible})$
as a function of the axion proper decay length. This model-independent
formulation does not assume a specific relation among the production
rate, decay width, and underlying couplings, enabling direct comparisons
with other searches for long-lived particles from rare kaon decays. 

Although our constraints do not exceed the strongest existing beam-dump
bounds, atmospheric kaon decays provide a distinct experimental probe,
characterized by a broad energy spectrum, an extended production region,
and propagation baselines different from those of accelerator-based
experiments. Large underground neutrino detectors therefore offer
independent and complementary sensitivity to long-lived particles
produced in atmospheric meson decays.

Future experiments, such as
Hyper-Kamiokande~\cite{Hyper-Kamiokande:2018ofw} and
IceCube-Gen2~\cite{IceCube-Gen2:2020qha}, could extend this sensitivity
through their larger detector volumes and exposures, providing new
opportunities to probe axions and other feebly interacting, long-lived
particles.

\section{Acknowledgments}
I am grateful to Vedran Brdar, Carlos Argüelles, and Gongjun Choi
for helpful discussions and valuable comments. I also thank Kevin Kelly
and Anatoli Fedynitch for useful conversations, and Mary Hall Reno,
Diksha Garg, Laksha Das, and Sumit Biswas for correspondence.

\appendix

\bibliography{apssamp}

@PREAMBLE{
 "\providecommand{\noopsort}[1]{}" 
 # "\providecommand{\singleletter}[1]{#1}%" 
}

@article{Witten:1984dg,
    author = "Witten, Edward",
    title = "{Some Properties of O(32) Superstrings}",
    journal = "Nucl. Phys. B",
    volume = "268",
    pages = "79--112",
    year = "1986",
    doi = "10.1016/0550-3213(86)90155-5"
}

@article{Choi:2021aze,
    author = "Choi, Gongjun and Lin, Weikang and Visinelli, Luca and Yanagida, Tsutomu T.",
    title = "{Cosmic birefringence and electroweak axion dark energy}",
    eprint = "2106.12602",
    archivePrefix = "arXiv",
    primaryClass = "hep-ph",
    doi = "10.1103/PhysRevD.104.L101302",
    journal = "Phys. Rev. D",
    volume = "104",
    number = "10",
    pages = "L101302",
    year = "2021"
}

@article{Ibe:2018ffn,
    author = "Ibe, Masahito and Yamazaki, Masahito and Yanagida, Tsutomu T.",
    title = "{Quintessence Axion Revisited in Light of Swampland Conjectures}",
    eprint = "1811.04664",
    archivePrefix = "arXiv",
    primaryClass = "hep-th",
    reportNumber = "IPMU-18-0183",
    doi = "10.1088/1361-6382/ab5197",
    journal = "Class. Quant. Grav.",
    volume = "36",
    number = "23",
    pages = "235020",
    year = "2019"
}

@article{Nomura:2000yk,
    author = "Nomura, Yasunori and Watari, T. and Yanagida, T.",
    title = "{Quintessence axion potential induced by electroweak instanton effects}",
    eprint = "hep-ph/0004182",
    archivePrefix = "arXiv",
    reportNumber = "UT-883",
    doi = "10.1016/S0370-2693(00)00605-5",
    journal = "Phys. Lett. B",
    volume = "484",
    pages = "103--111",
    year = "2000"
}

@article{Choi:2019jck,
    author = "Choi, Gongjun and Suzuki, Motoo and Yanagida, Tsutomu T.",
    title = "{Quintessence axion dark energy and a solution to the hubble tension}",
    eprint = "1910.00459",
    archivePrefix = "arXiv",
    primaryClass = "hep-ph",
    doi = "10.1016/j.physletb.2020.135408",
    journal = "Phys. Lett. B",
    volume = "805",
    pages = "135408",
    year = "2020"
}

@article{Kaloper:2008fb,
    author = "Kaloper, Nemanja and Sorbo, Lorenzo",
    title = "{Where in the String Landscape is Quintessence}",
    journal = "Phys. Rev. D",
    volume = "79",
    pages = "043528",
    year = "2009",
    eprint = "0810.5346",
    archivePrefix = "arXiv",
    primaryClass = "hep-th",
    doi = "10.1103/PhysRevD.79.043528"
}

@article{Kaloper:2005aj,
    author = "Kaloper, Nemanja and Sorbo, Lorenzo",
    title = "{Of pNGB Quintessence}",
    journal = "JCAP",
    volume = "05",
    pages = "010",
    year = "2006",
    eprint = "astro-ph/0511543",
    archivePrefix = "arXiv",
    primaryClass = "astro-ph",
    doi = "10.1088/1475-7516/2006/05/010"
}

@article{Frieman:1995pm,
    author = "Frieman, Joshua A. and Hill, Christopher T. and Stebbins, Albert and Waga, Ichiro",
    title = "{Cosmology with Ultralight Pseudo Nambu-Goldstone Bosons}",
    journal = "Phys. Rev. Lett.",
    volume = "75",
    pages = "2077--2080",
    year = "1995",
    eprint = "astro-ph/9505060",
    archivePrefix = "arXiv",
    doi = "10.1103/PhysRevLett.75.2077"
}

@article{Kim:2004rp,
    author = "Kim, Jihn E. and Nilles, Hans Peter and Peloso, Marco",
    title = "{Completing Natural Inflation}",
    journal = "JCAP",
    volume = "01",
    pages = "005",
    year = "2005",
    eprint = "hep-ph/0409138",
    archivePrefix = "arXiv",
    primaryClass = "hep-ph",
    doi = "10.1088/1475-7516/2005/01/005"
}

@article{Preskill:1982cy,
    author = "Preskill, John and Wise, Mark B. and Wilczek, Frank",
    title = "{Cosmology of the Invisible Axion}",
    journal = "Phys. Lett. B",
    volume = "120",
    pages = "127--132",
    year = "1983",
    doi = "10.1016/0370-2693(83)90637-8"
}

@article{Abbott:1982af,
    author = "Abbott, Laurence F. and Sikivie, Pierre",
    title = "{A Cosmological Bound on the Invisible Axion}",
    journal = "Phys. Lett. B",
    volume = "120",
    pages = "133--136",
    year = "1983",
    doi = "10.1016/0370-2693(83)90638-X"
}

@article{Dine:1982ah,
    author = "Dine, Michael and Fischler, Willy",
    title = "{The Not So Harmless Axion}",
    journal = "Phys. Lett. B",
    volume = "120",
    pages = "137--141",
    year = "1983",
    doi = "10.1016/0370-2693(83)90639-1"
}

@article{Freese:1990rb,
    author = "Freese, Katherine and Frieman, Joshua A. and Olinto, Angela V.",
    title = "{Natural Inflation with Pseudo - Nambu-Goldstone Bosons}",
    journal = "Phys. Rev. Lett.",
    volume = "65",
    pages = "3233--3236",
    year = "1990",
    doi = "10.1103/PhysRevLett.65.3233"
}

@article{Freese:2004un,
    author = "Freese, Katherine and Kinney, William H.",
    title = "{Natural Inflation: Consistency with Cosmic Microwave Background Observations}",
    journal = "Phys. Rev. D",
    volume = "70",
    pages = "083512",
    year = "2004",
    eprint = "hep-ph/0404012",
    archivePrefix = "arXiv",
    doi = "10.1103/PhysRevD.70.083512"
}

@article{Silverstein:2008sg,
    author = "Silverstein, Eva and Westphal, Alexander",
    title = "{Monodromy in the CMB: Gravity Waves and String Inflation}",
    journal = "Phys. Rev. D",
    volume = "78",
    pages = "106003",
    year = "2008",
    eprint = "0803.3085",
    archivePrefix = "arXiv",
    primaryClass = "hep-th",
    doi = "10.1103/PhysRevD.78.106003"
}

@article{Svrcek:2006yi,
    author = "Svrcek, Peter and Witten, Edward",
    title = "{Axions In String Theory}",
    journal = "JHEP",
    volume = "06",
    pages = "051",
    year = "2006",
    eprint = "hep-th/0605206",
    archivePrefix = "arXiv",
    primaryClass = "hep-th",
    doi = "10.1088/1126-6708/2006/06/051"
}

@article{Arvanitaki:2009fg,
    author = "Arvanitaki, Asimina and Dimopoulos, Savas and Dubovsky, Sergei and Kaloper, Nemanja and March-Russell, John",
    title = "{String Axiverse}",
    eprint = "0905.4720",
    archivePrefix = "arXiv",
    primaryClass = "hep-th",
    doi = "10.1103/PhysRevD.81.123530",
    journal = "Phys. Rev. D",
    volume = "81",
    pages = "123530",
    year = "2010"
}

@article{ParticleDataGroup:2024cfk,
    author = "Navas, S. and others",
    collaboration = "Particle Data Group",
    title = "{Review of particle physics}",
    doi = "10.1103/PhysRevD.110.030001",
    journal = "Phys. Rev. D",
    volume = "110",
    number = "3",
    pages = "030001",
    year = "2024"
}

@article{Fedynitch:2015zma,
    author = "Fedynitch, Anatoli and Engel, Ralph and Gaisser, Thomas K. and Riehn, Felix and Stanev, Todor",
    editor = "Berge, D. and de Roeck, A. and Mangano, M. and Pattison, B.",
    title = "{Calculation of conventional and prompt lepton fluxes at very high energy}",
    eprint = "1503.00544",
    archivePrefix = "arXiv",
    primaryClass = "hep-ph",
    doi = "10.1051/epjconf/20159908001",
    journal = "EPJ Web Conf.",
    volume = "99",
    pages = "08001",
    year = "2015"
}

@article{Fedynitch:2018cbl,
    author = "Fedynitch, Anatoli and Riehn, Felix and Engel, Ralph and Gaisser, Thomas K. and Stanev, Todor",
    title = "{Hadronic interaction model sibyll 2.3c and inclusive lepton fluxes}",
    eprint = "1806.04140",
    archivePrefix = "arXiv",
    primaryClass = "hep-ph",
    reportNumber = "DESY-18-110",
    doi = "10.1103/PhysRevD.100.103018",
    journal = "Phys. Rev. D",
    volume = "100",
    number = "10",
    pages = "103018",
    year = "2019"
}

@article{Arguelles:2019ziu,
    author = {Arg{\"u}elles, Carlos and Coloma, Pilar and Hern{\'a}ndez, Pilar and Mu{\~n}oz, V{\'\i}ctor},
    title = "{Searches for Atmospheric Long-Lived Particles}",
    eprint = "1910.12839",
    archivePrefix = "arXiv",
    primaryClass = "hep-ph",
    doi = "10.1007/JHEP02(2020)190",
    journal = "JHEP",
    volume = "02",
    pages = "190",
    year = "2020"
}

@article{Bauer:2021wjo,
    author = "Bauer, Martin and Neubert, Matthias and Renner, Sophie and Schnubel, Marvin and Thamm, Andrea",
    title = "{Consistent Treatment of Axions in the Weak Chiral Lagrangian}",
    eprint = "2102.13112",
    archivePrefix = "arXiv",
    primaryClass = "hep-ph",
    reportNumber = "IPPP/20-82, MITP/21-007, ZU-TH-01/21",
    doi = "10.1103/PhysRevLett.127.081803",
    journal = "Phys. Rev. Lett.",
    volume = "127",
    number = "8",
    pages = "081803",
    year = "2021"
}

@article{Super-Kamiokande:2017yvm,
    author = "Abe, K. and others",
    collaboration = "Super-Kamiokande",
    title = "{Atmospheric neutrino oscillation analysis with external constraints in Super-Kamiokande I-IV}",
    eprint = "1710.09126",
    archivePrefix = "arXiv",
    primaryClass = "hep-ex",
    doi = "10.1103/PhysRevD.97.072001",
    journal = "Phys. Rev. D",
    volume = "97",
    number = "7",
    pages = "072001",
    year = "2018"
}

@article{IceCube:2014rwe,
    author = "Aartsen, M. G. and others",
    collaboration = "IceCube",
    title = "{Atmospheric and astrophysical neutrinos above 1 TeV interacting in IceCube}",
    eprint = "1410.1749",
    archivePrefix = "arXiv",
    primaryClass = "astro-ph.HE",
    doi = "10.1103/PhysRevD.91.022001",
    journal = "Phys. Rev. D",
    volume = "91",
    number = "2",
    pages = "022001",
    year = "2015"
}

@article{Gondolo:1995fq,
    author = "Gondolo, P. and Ingelman, G. and Thunman, M.",
    title = "{Charm production and high-energy atmospheric muon and neutrino fluxes}",
    eprint = "hep-ph/9505417",
    archivePrefix = "arXiv",
    reportNumber = "TSL-ISV-95-0120, PAR-LPTHE-95-29",
    doi = "10.1016/0927-6505(96)00033-3",
    journal = "Astropart. Phys.",
    volume = "5",
    pages = "309--332",
    year = "1996"
}

@article{Ema:2023tjg,
    author = "Ema, Yohei and Liu, Zhen and Plestid, Ryan",
    title = "{Searching for axions with kaon decay at rest}",
    eprint = "2308.08589",
    archivePrefix = "arXiv",
    primaryClass = "hep-ph",
    reportNumber = "UMN-TH-4221/23, FTPI-MINN-23-13, CALT-TH/2023-028",
    doi = "10.1103/PhysRevD.109.L031702",
    journal = "Phys. Rev. D",
    volume = "109",
    number = "3",
    pages = "L031702",
    year = "2024"
}

@article{Kelly:2020dda,
    author = "Kelly, Kevin J. and Kumar, Soubhik and Liu, Zhen",
    title = "{Heavy axion opportunities at the DUNE near detector}",
    eprint = "2011.05995",
    archivePrefix = "arXiv",
    primaryClass = "hep-ph",
    reportNumber = "FERMILAB-PUB-20-581-T",
    doi = "10.1103/PhysRevD.103.095002",
    journal = "Phys. Rev. D",
    volume = "103",
    number = "9",
    pages = "095002",
    year = "2021"
}

@article{Chang:2018rso,
    author = "Chang, Jae Hyeok and Essig, Rouven and McDermott, Samuel D.",
    title = "{Supernova 1987A Constraints on Sub-GeV Dark Sectors, Millicharged Particles, the QCD Axion, and an Axion-like Particle}",
    eprint = "1803.00993",
    archivePrefix = "arXiv",
    primaryClass = "hep-ph",
    reportNumber = "YITP-SB-18-01, FERMILAB-PUB-17-432-T",
    doi = "10.1007/JHEP09(2018)051",
    journal = "JHEP",
    volume = "09",
    pages = "051",
    year = "2018"
}

@article{Ertas:2020xcc,
    author = "Ertas, Fatih and Kahlhoefer, Felix",
    title = "{On the interplay between astrophysical and laboratory probes of MeV-scale axion-like particles}",
    eprint = "2004.01193",
    archivePrefix = "arXiv",
    primaryClass = "hep-ph",
    reportNumber = "TTK-20-08",
    doi = "10.1007/JHEP07(2020)050",
    journal = "JHEP",
    volume = "07",
    pages = "050",
    year = "2020"
}

@article{Depta:2020wmr,
    author = "Depta, Paul Frederik and Hufnagel, Marco and Schmidt-Hoberg, Kai",
    title = "{Robust cosmological constraints on axion-like particles}",
    eprint = "2002.08370",
    archivePrefix = "arXiv",
    primaryClass = "hep-ph",
    reportNumber = "DESY-20-003, DESY 20-003",
    doi = "10.1088/1475-7516/2020/05/009",
    journal = "JCAP",
    volume = "05",
    pages = "009",
    year = "2020"
}

@article{BNL-E949:2009dza,
    author = "Artamonov, A. V. and others",
    collaboration = "BNL-E949",
    title = "{Study of the decay $K^+\to\pi^+\nu \bar\nu$ in the momentum region $140 < P_\pi < 199$ MeV/c}",
    eprint = "0903.0030",
    archivePrefix = "arXiv",
    primaryClass = "hep-ex",
    reportNumber = "BNL-81786-2008-JA, FERMILAB-PUB-09-007-CD-T, KEK-2008-44, TRIUMF-TRI-PP-08-26, UHEP-EX-08-004",
    doi = "10.1103/PhysRevD.79.092004",
    journal = "Phys. Rev. D",
    volume = "79",
    pages = "092004",
    year = "2009"
}

@article{NA62:2021zjw,
    author = "Cortina Gil, Eduardo and others",
    collaboration = "NA62",
    title = "{Measurement of the very rare K$^{+}${\textrightarrow}$ {\pi}^{+}\nu \overline{\nu} $ decay}",
    eprint = "2103.15389",
    archivePrefix = "arXiv",
    primaryClass = "hep-ex",
    doi = "10.1007/JHEP06(2021)093",
    journal = "JHEP",
    volume = "06",
    pages = "093",
    year = "2021"
}

@article{MicroBooNE:2021sov,
    author = "Abratenko, P. and others",
    collaboration = "MicroBooNE",
    title = "{Search for a Higgs Portal Scalar Decaying to Electron-Positron Pairs in the MicroBooNE Detector}",
    eprint = "2106.00568",
    archivePrefix = "arXiv",
    primaryClass = "hep-ex",
    reportNumber = "FERMILAB-MICROBOONE-NOTE-1092-PUB, FERMILAB-PUB-21-262-E",
    doi = "10.1103/PhysRevLett.127.151803",
    journal = "Phys. Rev. Lett.",
    volume = "127",
    number = "15",
    pages = "151803",
    year = "2021"
}

@article{CHARM:1985anb,
    author = "Bergsma, F. and others",
    collaboration = "CHARM",
    title = "{Search for Axion Like Particle Production in 400-{GeV} Proton - Copper Interactions}",
    reportNumber = "CERN-EP-85-38",
    doi = "10.1016/0370-2693(85)90400-9",
    journal = "Phys. Lett. B",
    volume = "157",
    pages = "458--462",
    year = "1985"
}

@article{Blumlein:1990ay,
    author = "Blumlein, J. and others",
    title = "{Limits on neutral light scalar and pseudoscalar particles in a proton beam dump experiment}",
    reportNumber = "PHE-90-03",
    doi = "10.1007/BF01548556",
    journal = "Z. Phys. C",
    volume = "51",
    pages = "341--350",
    year = "1991"
}

@article{Blinov:2021say,
    author = "Blinov, Nikita and Kowalczyk, Elizabeth and Wynne, Margaret",
    title = "{Axion-like particle searches at DarkQuest}",
    eprint = "2112.09814",
    archivePrefix = "arXiv",
    primaryClass = "hep-ph",
    reportNumber = "FERMILAB-PUB-21-749-V",
    doi = "10.1007/JHEP02(2022)036",
    journal = "JHEP",
    volume = "02",
    pages = "036",
    year = "2022"
}

@article{FASER:2018eoc,
    author = "Ariga, Akitaka and others",
    collaboration = "FASER",
    title = "{FASER{\textquoteright}s physics reach for long-lived particles}",
    eprint = "1811.12522",
    archivePrefix = "arXiv",
    primaryClass = "hep-ph",
    reportNumber = "UCI-TR-2018-19, KYUSHU-RCAPP-2018-06",
    doi = "10.1103/PhysRevD.99.095011",
    journal = "Phys. Rev. D",
    volume = "99",
    number = "9",
    pages = "095011",
    year = "2019"
}

@article{SHiP:2018xqw,
    author = "Ahdida, C. and others",
    collaboration = "SHiP",
    title = "{Sensitivity of the SHiP experiment to Heavy Neutral Leptons}",
    eprint = "1811.00930",
    archivePrefix = "arXiv",
    primaryClass = "hep-ph",
    doi = "10.1007/JHEP04(2019)077",
    journal = "JHEP",
    volume = "04",
    pages = "077",
    year = "2019"
}

@article{Ostapchenko:2024myl,
    author = "Ostapchenko, Sergey",
    title = "{QGSJET-III model of high energy hadronic interactions. II. Particle production and extensive air shower characteristics}",
    eprint = "2403.16106",
    archivePrefix = "arXiv",
    primaryClass = "hep-ph",
    doi = "10.1103/PhysRevD.109.094019",
    journal = "Phys. Rev. D",
    volume = "109",
    number = "9",
    pages = "094019",
    year = "2024"
}

@phdthesis{Fedynitch:2015kcn,
    author = "Fedynitch, Anatoli",
    title = "{Cascade equations and hadronic interactions at very high energies}",
    reportNumber = "CERN-THESIS-2015-371",
    doi = "10.5445/IR/1000055433",
    school = "KIT, Karlsruhe, Dept. Phys.",
    month = "11",
    year = "2015"
}

@inproceedings{Roesler:2000he,
    author = "Roesler, Stefan and Engel, Ralph and Ranft, Johannes",
    title = "{The Monte Carlo event generator DPMJET-III}",
    booktitle = "{International Conference on Advanced Monte Carlo for Radiation Physics, Particle Transport Simulation and Applications (MC 2000)}",
    eprint = "hep-ph/0012252",
    archivePrefix = "arXiv",
    reportNumber = "SLAC-PUB-8740",
    doi = "10.1007/978-3-642-18211-2_166",
    pages = "1033--1038",
    month = "12",
    year = "2000"
}

@article{Pierog:2013ria,
    author = "Pierog, T. and Karpenko, Iu. and Katzy, J. M. and Yatsenko, E. and Werner, K.",
    title = "{EPOS LHC: Test of collective hadronization with data measured at the CERN Large Hadron Collider}",
    eprint = "1306.0121",
    archivePrefix = "arXiv",
    primaryClass = "hep-ph",
    reportNumber = "DESY-13-125",
    doi = "10.1103/PhysRevC.92.034906",
    journal = "Phys. Rev. C",
    volume = "92",
    number = "3",
    pages = "034906",
    year = "2015"
}

@ARTICLE{2012PhRvD..86k4024F,
       author = {{Fedynitch}, Anatoli and {Becker Tjus}, Julia and {Desiati}, Paolo},
        title = "{Influence of hadronic interaction models and the cosmic ray spectrum on the high energy atmospheric muon and neutrino flux}",
      journal = {\prd},
         year = 2012,
        month = dec,
       volume = {86},
       number = {11},
          eid = {114024},
        pages = {114024},
          doi = {10.1103/PhysRevD.86.114024},
archivePrefix = {arXiv},
       eprint = {1206.6710},
 primaryClass = {astro-ph.HE},
       adsurl = {https://ui.adsabs.harvard.edu/abs/2012PhRvD..86k4024F}
}

@ARTICLE{2016PhRvL.116p1601G,
       author = {{Graner}, B. and {Chen}, Y. and {Lindahl}, E.~G. and {Heckel}, B.~R.},
        title = "{Reduced Limit on the Permanent Electric Dipole Moment of <mml:mmultiscripts>Hg 199 </mml:mmultiscripts>}",
      journal = {\prl},
         year = 2016,
        month = apr,
       volume = {116},
       number = {16},
          eid = {161601},
        pages = {161601},
          doi = {10.1103/PhysRevLett.116.161601},
archivePrefix = {arXiv},
       eprint = {1601.04339},
 primaryClass = {physics.atom-ph},
       adsurl = {https://ui.adsabs.harvard.edu/abs/2016PhRvL.116p1601G}
}

@article{Abel:2020pzs,
    author = "Abel, C. and others",
    title = "{Measurement of the Permanent Electric Dipole Moment of the Neutron}",
    eprint = "2001.11966",
    archivePrefix = "arXiv",
    primaryClass = "hep-ex",
    doi = "10.1103/PhysRevLett.124.081803",
    journal = "Phys. Rev. Lett.",
    volume = "124",
    number = "8",
    pages = "081803",
    year = "2020"
}

@article{PhysRevLett.38.1440,
  title = {$\mathrm{CP}$ Conservation in the Presence of Pseudoparticles},
  author = {Peccei, R. D. and Quinn, Helen R.},
  journal = {Phys. Rev. Lett.},
  volume = {38},
  issue = {25},
  pages = {1440--1443},
  numpages = {0},
  year = {1977},
  month = {Jun},
  publisher = {American Physical Society},
  doi = {10.1103/PhysRevLett.38.1440},
  url = {https://link.aps.org/doi/10.1103/PhysRevLett.38.1440}
}

@article{PhysRevD.16.1791,
  title = {Constraints imposed by $\mathrm{CP}$ conservation in the presence of pseudoparticles},
  author = {Peccei, R. D. and Quinn, Helen R.},
  journal = {Phys. Rev. D},
  volume = {16},
  issue = {6},
  pages = {1791--1797},
  numpages = {0},
  year = {1977},
  month = {Sep},
  publisher = {American Physical Society},
  doi = {10.1103/PhysRevD.16.1791},
  url = {https://link.aps.org/doi/10.1103/PhysRevD.16.1791}
}

@article{PhysRevLett.40.223,
  title = {A New Light Boson?},
  author = {Weinberg, Steven},
  journal = {Phys. Rev. Lett.},
  volume = {40},
  issue = {4},
  pages = {223--226},
  numpages = {0},
  year = {1978},
  month = {Jan},
  publisher = {American Physical Society},
  doi = {10.1103/PhysRevLett.40.223},
  url = {https://link.aps.org/doi/10.1103/PhysRevLett.40.223}
}

@article{PhysRevLett.40.279,
  title = {Problem of Strong $P$ and $T$ Invariance in the Presence of Instantons},
  author = {Wilczek, F.},
  journal = {Phys. Rev. Lett.},
  volume = {40},
  issue = {5},
  pages = {279--282},
  numpages = {0},
  year = {1978},
  month = {Jan},
  publisher = {American Physical Society},
  doi = {10.1103/PhysRevLett.40.279},
  url = {https://link.aps.org/doi/10.1103/PhysRevLett.40.279}
}

@article{Cheung:2022umw,
    author = "Cheung, Kingman and Kuo, Jui-Lin and Tseng, Po-Yan and Wang, Zeren Simon",
    title = "{Atmospheric axionlike particles at Super-Kamiokande}",
    eprint = "2208.05111",
    archivePrefix = "arXiv",
    primaryClass = "hep-ph",
    doi = "10.1103/PhysRevD.106.095029",
    journal = "Phys. Rev. D",
    volume = "106",
    number = "9",
    pages = "095029",
    year = "2022"
}

@inproceedings{Essig:2013lka,
    author = "Essig, Rouven and others",
    title = "{Working Group Report: New Light Weakly Coupled Particles}",
    booktitle = "{Snowmass 2013}: {Snowmass on the Mississippi}",
    eprint = "1311.0029",
    archivePrefix = "arXiv",
    primaryClass = "hep-ph",
    reportNumber = "YITP-SB-36, FERMILAB-CONF-13-653",
    month = "10",
    year = "2013"
}

@article{Hook:2019qoh,
    author = "Hook, Anson and Kumar, Soubhik and Liu, Zhen and Sundrum, Raman",
    title = "{High Quality QCD Axion and the LHC}",
    eprint = "1911.12364",
    archivePrefix = "arXiv",
    primaryClass = "hep-ph",
    reportNumber = "UMD-PP-019-07",
    doi = "10.1103/PhysRevLett.124.221801",
    journal = "Phys. Rev. Lett.",
    volume = "124",
    number = "22",
    pages = "221801",
    year = "2020"
}

@article{Goudzovski:2022vbt,
    author = "Goudzovski, Evgueni and others",
    title = "{New physics searches at kaon and hyperon factories}",
    eprint = "2201.07805",
    archivePrefix = "arXiv",
    primaryClass = "hep-ph",
    reportNumber = "FERMILAB-PUB-22-057-T",
    doi = "10.1088/1361-6633/ac9cee",
    journal = "Rept. Prog. Phys.",
    volume = "86",
    number = "1",
    pages = "016201",
    year = "2023"
}

@article{NA62:2023olg,
    author = "Cortina Gil, Eduardo and others",
    collaboration = "NA62",
    title = "{Measurement of the K+{\textrightarrow}{\ensuremath{\pi}}+{\ensuremath{\gamma}}{\ensuremath{\gamma}} decay}",
    eprint = "2311.01837",
    archivePrefix = "arXiv",
    primaryClass = "hep-ex",
    reportNumber = "CERN-EP-2023-247",
    doi = "10.1016/j.physletb.2024.138513",
    journal = "Phys. Lett. B",
    volume = "850",
    pages = "138513",
    year = "2024"
}

@article{NA62:2025upx,
    author = "Cortina Gil, Eduardo and others",
    collaboration = "NA62",
    title = "{Searches for hidden sectors using $K^+\to\pi^+X$ decays}",
    eprint = "2507.17286",
    archivePrefix = "arXiv",
    primaryClass = "hep-ex",
    reportNumber = "CERN-EP-2025-167",
    doi = "10.1007/JHEP11(2025)143",
    journal = "JHEP",
    volume = "11",
    pages = "143",
    year = "2025"
}

@article{ArgoNeuT:2022mrm,
    author = "Acciarri, R. and others",
    collaboration = "ArgoNeuT",
    title = "{First Constraints on Heavy QCD Axions with a Liquid Argon Time Projection Chamber Using the ArgoNeuT Experiment}",
    eprint = "2207.08448",
    archivePrefix = "arXiv",
    primaryClass = "hep-ex",
    reportNumber = "FERMILAB-PUB-22-527-ND-T, UMN-TH-4128/22, FTPI-MINN-22/19",
    doi = "10.1103/PhysRevLett.130.221802",
    journal = "Phys. Rev. Lett.",
    volume = "130",
    number = "22",
    pages = "221802",
    year = "2023"
}

@article{BDX:2016akw,
    author = "Battaglieri, M. and others",
    collaboration = "BDX",
    title = "{Dark Matter Search in a Beam-Dump eXperiment (BDX) at Jefferson Lab}",
    eprint = "1607.01390",
    archivePrefix = "arXiv",
    primaryClass = "hep-ex",
    reportNumber = "FERMILAB-TM-2630-PPD",
    month = "7",
    year = "2016"
}

@article{PIONEER:2022yag,
    author = "Altmannshofer, W. and others",
    collaboration = "PIONEER",
    title = "{PIONEER: Studies of Rare Pion Decays}",
    eprint = "2203.01981",
    archivePrefix = "arXiv",
    primaryClass = "hep-ex",
    month = "3",
    year = "2022"
}

@article{PIENU:2021clt,
    author = "Aguilar-Arevalo, A. and others",
    collaboration = "PIENU",
    title = "{Search for three body pion decays ${\pi}^+{\to}l^+{\nu}X$}",
    eprint = "2101.07381",
    archivePrefix = "arXiv",
    primaryClass = "hep-ex",
    doi = "10.1103/PhysRevD.103.052006",
    journal = "Phys. Rev. D",
    volume = "103",
    number = "5",
    pages = "052006",
    year = "2021"
}

@article{LDMX:2018cma,
    author = "{\r{A}}kesson, Torsten and others",
    collaboration = "LDMX",
    title = "{Light Dark Matter eXperiment (LDMX)}",
    eprint = "1808.05219",
    archivePrefix = "arXiv",
    primaryClass = "hep-ex",
    reportNumber = "FERMILAB-PUB-18-324-A, SLAC-PUB-17303",
    month = "8",
    year = "2018"
}

@article{Ema:2025bww,
    author = "Ema, Yohei and Fox, Patrick J. and Hostert, Matheus and Menzo, Tony and Pospelov, Maxim and Ray, Anupam and Zupan, Jure",
    title = "{Long-lived axionlike particles from tau decays}",
    eprint = "2507.15271",
    archivePrefix = "arXiv",
    primaryClass = "hep-ph",
    reportNumber = "CERN-TH-2025-123, FERMILAB-PUB-25-0408-T, N3AS-25-011",
    doi = "10.1103/51gx-m32t",
    journal = "Phys. Rev. D",
    volume = "112",
    number = "11",
    pages = "115028",
    year = "2025"
}

@article{Dolan:2017osp,
    author = "Dolan, Matthew J. and Ferber, Torben and Hearty, Christopher and Kahlhoefer, Felix and Schmidt-Hoberg, Kai",
    title = "{Revised constraints and Belle II sensitivity for visible and invisible axion-like particles}",
    eprint = "1709.00009",
    archivePrefix = "arXiv",
    primaryClass = "hep-ph",
    reportNumber = "DESY-17-127",
    doi = "10.1007/JHEP12(2017)094",
    journal = "JHEP",
    volume = "12",
    pages = "094",
    year = "2017",
    note = "[Erratum: JHEP 03, 190 (2021)]"
}

@article{Coloma:2023oxx,
    author = "Coloma, Pilar and Mart{\'\i}n-Albo, Justo and Urrea, Salvador",
    title = "{Discovering long-lived particles at DUNE}",
    eprint = "2309.06492",
    archivePrefix = "arXiv",
    primaryClass = "hep-ph",
    reportNumber = "IFT-UAM/CSIC-23-111, IFIC/23-40, FTUV-23-0823.4331",
    doi = "10.1103/PhysRevD.109.035013",
    journal = "Phys. Rev. D",
    volume = "109",
    number = "3",
    pages = "035013",
    year = "2024"
}

@article{NA64:2020qwq,
    author = "Banerjee, D. and others",
    collaboration = "NA64",
    title = "{Search for Axionlike and Scalar Particles with the NA64 Experiment}",
    eprint = "2005.02710",
    archivePrefix = "arXiv",
    primaryClass = "hep-ex",
    reportNumber = "CERN-EP-2020-068",
    doi = "10.1103/PhysRevLett.125.081801",
    journal = "Phys. Rev. Lett.",
    volume = "125",
    number = "8",
    pages = "081801",
    year = "2020"
}

@article{Hyper-Kamiokande:2018ofw,
    author = "Abe, K. and others",
    collaboration = "Hyper-Kamiokande",
    title = "{Hyper-Kamiokande Design Report}",
    eprint = "1805.04163",
    archivePrefix = "arXiv",
    primaryClass = "physics.ins-det",
    month = "5",
    year = "2018"
}

@article{IceCube-Gen2:2020qha,
    author = "Aartsen, M. G. and others",
    collaboration = "IceCube-Gen2",
    title = "{IceCube-Gen2: the window to the extreme Universe}",
    eprint = "2008.04323",
    archivePrefix = "arXiv",
    primaryClass = "astro-ph.HE",
    doi = "10.1088/1361-6471/abbd48",
    journal = "J. Phys. G",
    volume = "48",
    number = "6",
    pages = "060501",
    year = "2021"
}

@article{Dobrich:2015jyk,
    author = {D{\"o}brich, Babette and Jaeckel, Joerg and Kahlhoefer, Felix and Ringwald, Andreas and Schmidt-Hoberg, Kai},
    title = "{ALPtraum: ALP production in proton beam dump experiments}",
    eprint = "1512.03069",
    archivePrefix = "arXiv",
    primaryClass = "hep-ph",
    reportNumber = "CERN-PH-TH-2015-293, DESY-15-237",
    doi = "10.1007/JHEP02(2016)018",
    journal = "JHEP",
    volume = "02",
    pages = "018",
    year = "2016"
}

@article{Gavela:2019cmq,
    author = "Gavela, M. B. and No, J. M. and Sanz, V. and de Troc{\'o}niz, J. F.",
    title = "{Nonresonant Searches for Axionlike Particles at the LHC}",
    eprint = "1905.12953",
    archivePrefix = "arXiv",
    primaryClass = "hep-ph",
    doi = "10.1103/PhysRevLett.124.051802",
    journal = "Phys. Rev. Lett.",
    volume = "124",
    number = "5",
    pages = "051802",
    year = "2020"
}

@article{Eberhart:2025lyu,
    author = "Eberhart, Alexander and Fedele, Marco and Kahlhoefer, Felix and Ravensburg, Eike and Ziegler, Robert",
    title = "{Leptophilic ALPs in laboratory experiments}",
    eprint = "2504.05873",
    archivePrefix = "arXiv",
    primaryClass = "hep-ph",
    reportNumber = "TTP25-011, P3H-25-026, MITP-25-027",
    doi = "10.1007/JHEP12(2025)055",
    journal = "JHEP",
    volume = "12",
    pages = "055",
    year = "2025"
}

@article{Choi:2020rgn,
    author = "Choi, Kiwoon and Im, Sang Hui and Shin, Chang Sub",
    title = "{Recent Progress in the Physics of Axions and Axion-Like Particles}",
    eprint = "2012.05029",
    archivePrefix = "arXiv",
    primaryClass = "hep-ph",
    reportNumber = "CTPU-PTC-20-28",
    doi = "10.1146/annurev-nucl-120720-031147",
    journal = "Ann. Rev. Nucl. Part. Sci.",
    volume = "71",
    pages = "225--252",
    year = "2021"
}

@article{Graham:2015ouw,
    author = "Graham, Peter W. and Irastorza, Igor G. and Lamoreaux, Steven K. and Lindner, Axel and van Bibber, Karl A.",
    title = "{Experimental Searches for the Axion and Axion-Like Particles}",
    eprint = "1602.00039",
    archivePrefix = "arXiv",
    primaryClass = "hep-ex",
    doi = "10.1146/annurev-nucl-102014-022120",
    journal = "Ann. Rev. Nucl. Part. Sci.",
    volume = "65",
    pages = "485--514",
    year = "2015"
}

@inproceedings{Marsh:2017hbv,
    author = "Marsh, David J. E.",
    title = "{Axions and ALPs: a very short introduction}",
    booktitle = "{13th Patras Workshop on Axions, WIMPs and WISPs}",
    eprint = "1712.03018",
    archivePrefix = "arXiv",
    primaryClass = "hep-ph",
    doi = "10.3204/DESY-PROC-2017-02/marsh_david",
    pages = "59--74",
    year = "2018"
}

@article{https://doi.org/10.1029/2002JA009430,
author = {Picone, J. M. and Hedin, A. E. and Drob, D. P. and Aikin, A. C.},
title = {NRLMSISE-00 empirical model of the atmosphere: Statistical comparisons and scientific issues},
journal = {Journal of Geophysical Research: Space Physics},
volume = {107},
number = {A12},
pages = {SIA 15-1-SIA 15-16},
doi = {https://doi.org/10.1029/2002JA009430},
url = {https://agupubs.onlinelibrary.wiley.com/doi/abs/10.1029/2002JA009430},
year = {2002}
}

@article{Gaisser:2011klf,
    author = "Gaisser, Thomas K.",
    title = "{Spectrum of cosmic-ray nucleons, kaon production, and the atmospheric muon charge ratio}",
    eprint = "1111.6675",
    archivePrefix = "arXiv",
    primaryClass = "astro-ph.HE",
    doi = "10.1016/j.astropartphys.2012.02.010",
    journal = "Astropart. Phys.",
    volume = "35",
    pages = "801--806",
    year = "2012"
}

\end{document}